\RequirePackage{fixltx2e}
\RequirePackage{atbegshi}
\documentclass[aps,prl,letterpaper,floatfix,twocolumn,superscriptaddress,amsmath,amsfonts,amssymb,floatfix,preprintnumbers,citeautoscript,10pt]{revtex4-1}

\usepackage[english]{babel}

\usepackage[ansinew]{inputenc}

\usepackage[T1]{fontenc}

\usepackage[space-before-unit,range-units = repeat]{siunitx}

\usepackage{graphicx}

\usepackage[colorlinks,plainpages=false,linkcolor=black,urlcolor=blue,citecolor=blue,pdfpagemode=UseNone,pdfstartview=FitH]{hyperref} 

\newcommand{\EF}{\textit{E}\textsubscript{F}}

\newcommand{\GM}{$\Gamma$}
\newcommand{\GMO}{$\overline{\Gamma}$}
\newcommand{\MO}{$\overline{\text{M}}$}

\newcommand{\TC}{\textit{T}\textsubscript{c}}

\newcommand{\kx}{\textit{k\textsubscript{x}}}
\newcommand{\ky}{\textit{k\textsubscript{y}}}
\newcommand{\kz}{\textit{k\textsubscript{z}}}
\newcommand{\kf}{\textit{k\textsubscript{F}}}

\newcommand{\EB}{E_{\textrm{bind}}}

\newcommand{\boldk}{\boldsymbol{k}}

\newcommand{\tsub}[1]{\textsubscript{#1}}
\newcommand{\tsup}[1]{\textsuperscript{#1}}

\begin{document}
\title{Unusual band renormalization in the simplest iron based superconductor}
\author{J.\, Maletz}\author{V.\,B.\,Zabolotnyy}\author{D.\,V.\,Evtushinsky}\author{S.\,Thirupathaiah}\author{A.\,U.\,B.\,Wolter}\author{L.\,Harnagea}
\affiliation{Institute for Solid State Research, IFW Dresden, P.\,O.\,Box 270116, D-01171 Dresden, Germany}
\author{A.\,N.Yaresko}
\affiliation{Max-Planck-Institute for Solid State Research, Heisenbergstrasse 1, D-70569 Stuttgart, Germany}
\author{A.\,N.\,Vasiliev}
\affiliation{Department of Low Temperature Physics, Moscow State University, Moscow 119991, Russia}
\author{D.\,A.\,Chareev}
\affiliation{Institute of Experimental Mineralogy, Chernogolovka, Moscow Region, 142432, Russia}
\author{E.\,D.\,L.\,Rienks}
\affiliation{Helmholtz-Zentrum Berlin,
Albert-Einstein-Strasse 15, 12489 Berlin, Germany}
\author{B.\,B\"{u}chner}
\affiliation{Institute for Solid State Research, IFW Dresden, P.\,O.\,Box 270116, D-01171 Dresden, Germany}
\affiliation{Institut f\"ur Festk\"orperphysik, Technische Universit\"at Dresden, D-01171 Dresden, Germany}
\author{S.\,V.\,Borisenko}
\affiliation{Institute for Solid State Research, IFW Dresden, P.\,O.\,Box 270116, D-01171 Dresden, Germany}
\date{\today}

\begin{abstract}
\noindent The electronic structure of the iron chalcogenide superconductor FeSe\textsubscript{\textit{1-x}} was investigated by high-resolution angle-resolved photoemission spectroscopy (ARPES). The results were compared to DFT calculations showing some significant differences between the experimental electronic structure of FeSe\textsubscript{\textit{1-x}}, DFT calculations and existing data on FeSe\tsub{\textit{x}}Te\tsub{1-\textit{x}}. The bands undergo a pronounced orbital dependent renormalization, different from what was observed for FeSe\tsub{\textit{x}}Te\tsub{1-\textit{x}} and any other pnictides.
\end{abstract}

\maketitle
\begin{figure}[ht]
   \centering
      \includegraphics{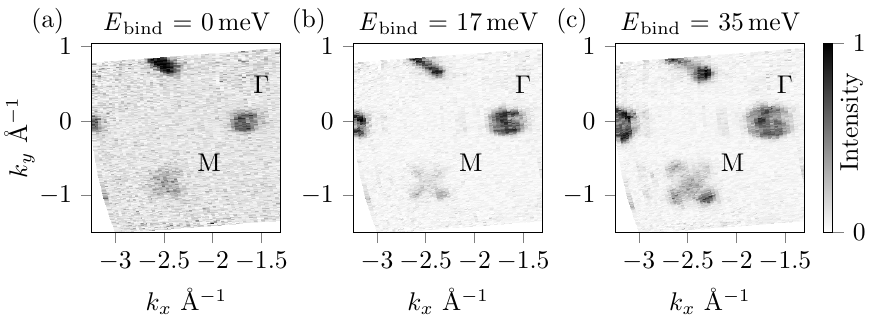}
		\includegraphics{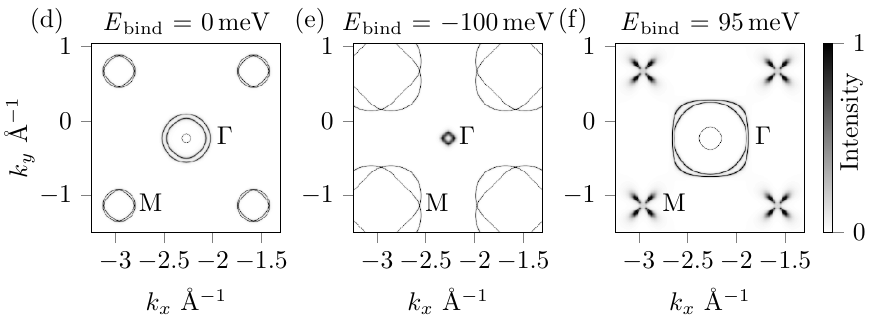}
	   \caption[Titel des Bildes]{\kx-\ky\ intensity map taken at \EF\ (a), at $\EB = \SI{17}{\milli\eV}$ (b) and at $\EB = \SI{35}{\milli\eV}$ (c) as measured by ARPES at $T = \SI{40}{\kelvin}$ with photon energy $h\nu = \SI{80}{\eV}$ and linear horizontal (LH) polarization. Spectra were energy integrated over an intervall of $\pm\SI{5}{\milli\eV}$. Fermi surface from DFT calculations for \EF\ (d), $\EB = \SI{-100}{\milli\eV}$ (e) and $\EB = \SI{95}{\milli\eV}$ (f).}
	\label{fig:FSMap}
\end{figure}
\begin{figure*}[tb]
\centering
	\includegraphics{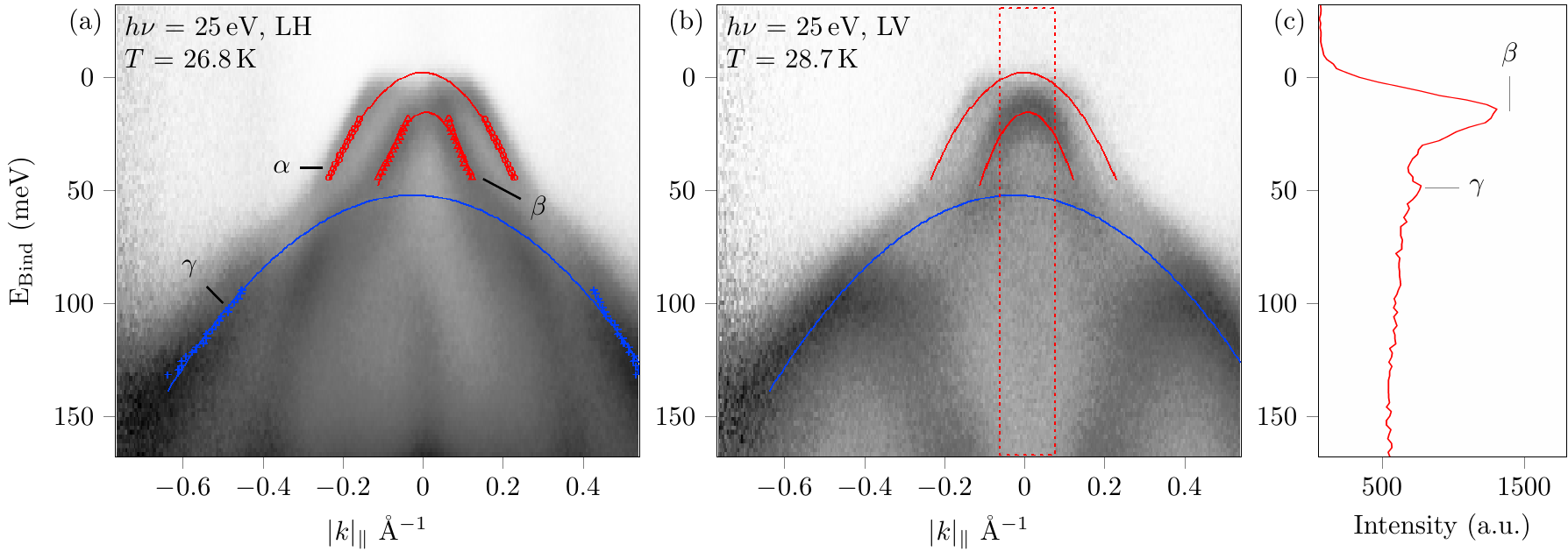}
	\includegraphics{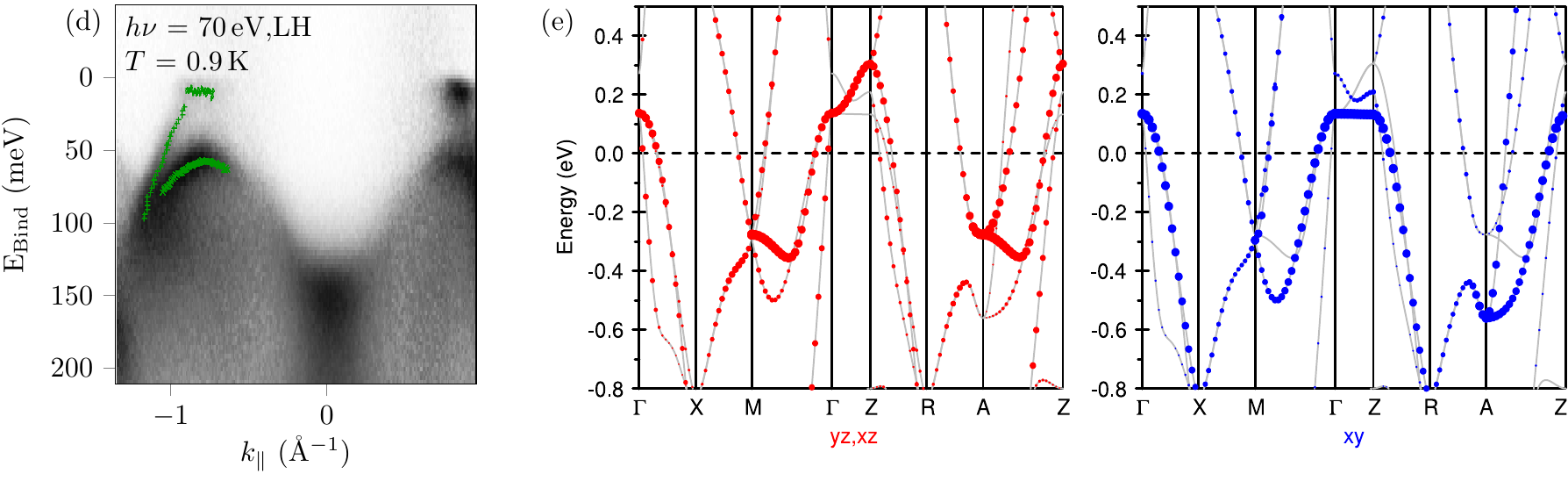}
	\caption{(a) Energy momentum cut taken at \SI{26.8}{\K} in \GM-X direction with $h\nu = \SI{25}{\eV}$ and linear horizontal (LH) polarization. The markers are the band dispersions derived from momentum distribution curves. The solid lines represent the expected band positions from fitting the obtained band positions. (b)  Energy momentum cut taken at \SI{28.7}{\K} in \GM-X direction with $h\nu = \SI{25}{\eV}$ and linear vertical (LV) polarization. (c) Integrated energy distribution curve from the dashed, boxed area in panel (b). The marked peaks correspond to the tops of the $\beta$- and $\gamma$-bands. (d) Energy momentum cut taken at \SI{0.9}{\K} in M-X direction. The markers represent the band dispersions derived from momentum and energy distribution curves. (e) LDA-LMTO calculations of FeSe, including orbital character.}
	\label{fig:GammaPointBands}
\end{figure*}

It is well established, that the iron arsenic/selenium layers, common for all the recently found and investigated iron based superconductors, are responsible for the superconductivity in these compounds\cite{Kamihara2006}. The binary ``11'' family of FeSe\textsubscript{\textit{1-x}} and FeSe\textsubscript{\textit{1-x}} offers the possibility, to investigate systems consiting of just these layers without the intermediate layers which are present in the ``111'', ``122'' and ``1111'' families. This simplest iron based superconductor may therefore yield valuable information about the origin of superconductivity in the iron pnictides/chalcogenides.
The FeSe\tsub{\textit{x}}Te\tsub{1-\textit{x}} ``11'' system exhibits long range antiferromagnetic (AFM) order for its end member FeTe which is supressed for $x > 0.1$, whereas a short range antiferromagnetic order appears for the intermediate range $0.1 < x <0.45$\cite{Fang2008,Khasanov2009}. Superconductivity is observed for $x > $ \numrange{0.2}{0.45}\cite{Yeh2008,Khasanov2009}, thus coexisting with the AFM also for the highest \TC\ composition of FeSe\tsub{0.4}Te\tsub{0.6}\cite{Yeh2008}. The end member FeSe has a \TC\ of $\sim \SI{8}{\K}$\cite{Hsu2008}. Recently, in a study on single layer FeSe grown on a SrTiO\tsub{3} substrate, superconductivity with a \TC\ of up to \SI{55}{\K}\cite{Wang2012}. Further studies on this system found an onset of superconductivity at up to \SI{65}{\K}\cite{He2012}.
There are also several angle-resolved photoemission spectroscopy (ARPES) studies on FeSe\tsub{\textit{x}}Te\tsub{1-\textit{x}} compounds, yielding interesting and in part contradicting results. Depending on \textit{x}, the position of the bands at \GMO\ point with respect to \EF\ may vary. For $x = 0.34$, all three low energy bands are crossing the Fermi level\cite{Chen2010}; for $x = $\numrange{0.3}{0.45} only two of them are crossing \EF\cite{Lubashevsky2012,Tamai2010,Miao2012,Nakayama2010}. For the monolayer FeSe, no Fermi surface in the zone center was observed at all\cite{Liu2012}. Additionally, there is no agreement concerning the renormalization of the bands, both in size and affected bands. These results show, that the electronic structure is highly dependent on the amount of tellurium and that a comparison with the calculated band structures for bulk FeSe and FeTe may yield rather large uncertaincies.
This issue can be adressed by investigating FeSe\textsubscript{\textit{1-x}} single crystals, as this allows to make more precise quantitative statements about the deviations from the DFT calculations.\\
Here we report on ARPES studies of FeSe$_{0.96}$ single crystals. The samples were grown using the KCl/AlCl\tsub{3} flux method, characterized with x-ray diffraction and EDX and investigated by low-temperature specific heat measurements showing a \TC\ of \SI{8.11}{\kelvin}\cite{Lin2011,Chareev2013}. ARPES measurements were performed at BESSY2 synchrotron facility. Samples were mounted on the cryo-manipulator of the 1\tsup{3}-ARPES station and cleaved at a temperature of $T \sim \SI{40}{\kelvin}$ in ultra high vacuum with a base pressure of $\sim \SI{e-11}{\milli\bar}$. Spectra were taken using excitation energies ranging from \SIrange{20}{120}{\eV} and temperatures down to \SI{890}{\milli\kelvin}. The overall energy and angular resolutions were $\Delta\text{E} = \SI{4}{\milli\eV}$ and $\Delta\eta = \ang{0.2}$ respectively.\\
\autoref{fig:FSMap} shows intensity maps of the $k_x$-$k_y$ planes at the Fermi energy (\EF), at $\EB = \SI{17}{\milli\eV}$ and at $\EB = \SI{35}{\milli\eV}$ (integrated over an interval of $\pm \SI{5}{\milli\eV}$) measured by ARPES as well as the Fermi surface contours for \EF, $\EB = \SI{-100}{\milli\eV}$ and $\EB = \SI{95}{\milli\eV}$ from DFT calculations.
The maps show two features, a circular hole-like Fermi surface around \GMO\ point (the features diameter grows with increasing binding energy) and a small electron-like Fermi surface around \MO\ point, which vanishes for higher binding energies and is replaced by a propeller shaped feature originating from two hole-like bands closing just beneath \EF\ at the \MO\ point. Comparing spectra and contours predicted by DFT, differences in both size and number of the Fermi surface sheets can be found. While from the calculations three Fermi surface sheets are expected around \GM\ point, the ARPES spectra show, that there is only one. This can be seen in in the energy momentum cuts in \autoref{fig:GammaPointBands} (a) and (b). The Fermi momentum (\kf) of the \GM-hole-pocket obtained from ARPES is \SI{0.05}{\per\angstrom} while calculations predict a \kf\ between \SI{0.25}{\per\angstrom} and \SI{0.3}{\per\angstrom}, which is 5 to 6 times larger than measured. The situation for the electron pockets located around the M-points of the Brillouin zone is similar. Here, the measured diameter is \SI{0.18}{\per\angstrom} while the calculated diameter was found to be \SI{0.37}{\per\angstrom}.
\autoref{fig:GammaPointBands} allows a closer look on the electronic structure at \GM- and M-points of the Brillouin zone. Panels (a) and (b) show energy-momentum cuts in \GM-X-direction taken with a photon energy $h\nu = \SI{25}{\eV}$ and two different polarizations. Panel (a) shows the cut corresponding to the linear horizontal (LH) polarization. Three bands can be identified, of which one, called $\alpha$ from hereon, crosses \EF\ to create a hole-like Fermi surface. From fitting the dispersion close to the Fermi energy, the top of this band is estimated to be at $\EB = \SI{-3}{\milli\eV}$. A second hole-like band, called $\beta$, also approaches \EF, though closing at $\EB = \SI{15}{\milli\eV}$, and thus not forming a Fermi surface. A third, flatter band, called $\gamma$, can be identified at higher binding energies, strongly losing intensity towards lower binding energies. From fitting the branches at higher binding energies, the top of this band is estimated to be at $\EB = \SI{52}{\milli\eV}$.
The fitted dispersions can also be seen in panel (b) which shows the same energy-momentum cut measured with linear vertical (LV) polarized light. In this polarization the top of the $\beta$-band is more prominent and matches the previously derived dispersion very well. Also, the top of the $\gamma$-band can be located in this polarization. This is shown in the integrated energy distribution curve (EDC) from the dashed area in panel (c). The EDC from the center of the Brillouin zone exhibits two peaks: one from the top of the $\beta$-band at $\EB = \SI{15}{\milli\eV}$ and the other one from the top of the $\gamma$-band at $\EB = \SI{49}{\milli\eV}$, both in very good agreement with the positions estimated from the previous fits.
\newline
In \autoref{fig:GammaPointBands} (d) an energy-momentum cut in M-X-direction is shown. Here one can clearly see shallow electron pockets at the M-points as well as two hole bands. The electron pockets close within \SIrange{5}{10}{\milli\eV} binding energy. One of the hole bands reaches up to the electron pockets, while the top of the other one is located around \SI{50}{\eV} binding energy. The hole band reaching up to the electron pocket is responsible for the propeller shaped features seen in \autoref{fig:FSMap} for higher binding energies. It remains unclear, whether the hole band reaches the Fermi level, but the calculations suggest that it remains below the electron pocket. For the \MO\ points, only one electron-like feature is observed. This is not suprising, as the calculations predict a degeneracy of the two electron pockets as well as some \kz-dispersion which would smear out the intensity of the two bands.\\
Comparing the energy-momentum cuts to the DFT calculations shown in \autoref{fig:GammaPointBands}(e), it can be seen that the bands are renormalized. The $\alpha$ and $\beta$ bands are renormalized by factors of $\sim 3$ and $\sim 3.7$ respectivly, while the $\gamma$ band is renormalized by a factor of $\sim 9$. This significantly differs from the previous studies on FeSe\tsub{\textit{x}}Te\tsub{1-\textit{x}}, where either a uniform renormalization of 2\cite{Xia2009,Nakayama2010} and 3.125\cite{Chen2010} had been observed, or a band selective renormalization of 1, 6, and 17\cite{Tamai2010} was reported, and is also unusual for iron based superconductors in general.\\
To match the experimental and calculated Fermi surfaces, the band renormalization is not sufficient. An additional shift has to be introduced. At \GM\ point, the three hole-like bands are shifted to higher binding energies, in a way, that only one of them remains crossing the Fermi level with a much smaller \kf\ than predicted. The same happens at the M-points, only that here the bands are shifted to lower binding energies, producing smaller electron pockets. 
At \GM\ point the shifts are \SI{0.09}{\eV}, \SI{0.065}{\eV} and \SI{0.045}{\eV} for the $\alpha$-, $\beta$- and $\gamma$-band respectivly. At M\ point, the electron band has to be shifted by \SI{-0.09}{\eV} while the hole bands approaching the electron pockets have to be shifted by \SIrange{-0.07}{-0.09}{\eV}. This behaviour, a k-dependent shift of the bandstructure, has already been observed in ARPES for other iron-pnictide compounds such as KFe\tsub{2}As\tsub{2}\cite{Yoshida2012} and LiFeAs\cite{Lee2012} and in quantum oscillation experiments for LaFePO\cite{Coldea2008} and seems to be a common feature for these compounds. \autoref{fig:FSMap} (e) shows the Fermi contours calculated for $\EB = \SI{-100}{\milli\eV}$ and reproduces the right size of the \GM\ Fermi surface sheet, while having way to large electron pockets at the M-points. Panel (f) shows the calculated contours for $\EB = \SI{95}{\milli\eV}$. Here, the propeller shaped features at the M-points from panel (b) ($\EB = \SI{17}{\milli\eV}$) are reproduced, while the calculated \GM\ Fermi surface is obviously too large.
\newline
\begin{figure}[htb]
\centering
	\includegraphics[width=\linewidth]{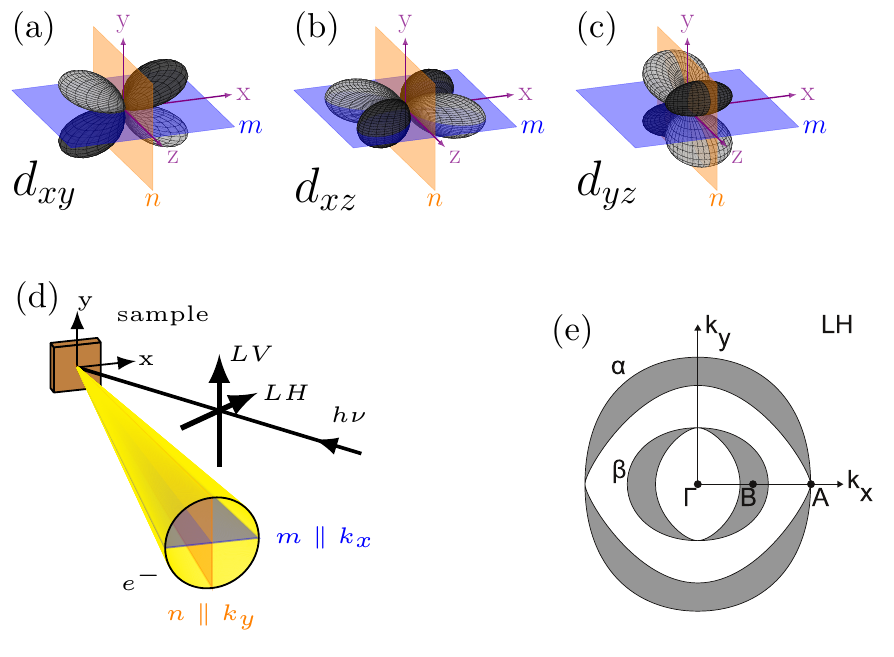}
	\caption{Orbital functions of: (a) $d_{xy}$, (b) $d_{xz}$ and (c) $d_{yz}$ orbitals with symmetry planes $m$ and $n$; (d) geometry of the angle resolved photoemission experiment; (e) sketch of the photoemission intensity around \GM\ point. The broader features correspond to a higher intensity observed in the experiment.}
	\label{fig:Fig-Symmetry-scaled}
\end{figure}
\begin{figure}[htb]
\centering
	\includegraphics[width=\linewidth]{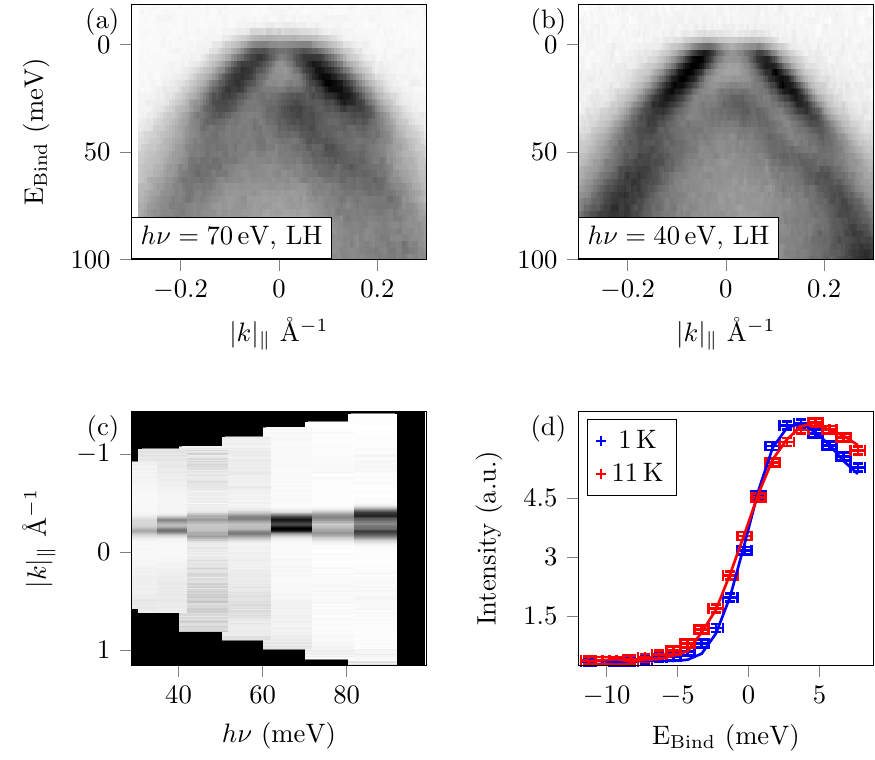}
	\caption{Energy momentum cuts taken at \GM-point with excitation energies of (a) \SI{70}{\eV} and (b) \SI{40}{\eV}. Panel (c) shows the evolution of the \GM\ Fermi surface sheet with excitation energy, showing a moderate cosine-like $k_{z}$ dispersion. (d) Energy distribution curves from \GM-point at \SI{1}{\kelvin} (superconducting state) and \SI{11}{\kelvin} (normal state). Markers represent the experimental data, solid lines were obtained by fitting to modelled data}
	\label{fig:kz_layout}
\end{figure}
From the collected data, one can also determine the orbital character of the bands. This provides the possibility to find correlations between band renormalization and orbital character. \autoref{fig:Fig-Symmetry-scaled} (e) shows a sketch of the intensity distribution around \GMO point as seen in \autoref{fig:FSMap} (b) and (c). There is almost zero intensity in point A.
Taking into account the symmetry of the experimental setup and photon polarization (compare \autoref{fig:Fig-Symmetry-scaled}), only $d_{xy}$ or $d_{yz}$ character of the bands explain this behaviour. Likewise, as the inner $\beta$ band shows intensity in the $m$ plane at point B, it cannot consist of $d_{xy}$ or $d_{yz}$ orbitals and therefore must be composed of $d_{xz}$ orbitals, as calculations offer only these three as possible candidates. As the $\alpha$ and $\beta$ bands show similar renormalization, we identify them with the predicted $d_{xz}/d_{yz}$ bands. Therefore, the $\gamma$ band has to be of $d_{xy}$ character. This is supported by the fact, that this band shows no \kz-dispersion, as predicted for the $d_{xy}$ band by calculations.\\
Additional temperature and energy dependent measurements were performed at \GM\ point. By using different excitation energies $h\nu$ in the range of \SIrange{33}{110}{\eV}, measurements for different \kz\ along the \GM-Z-direction were conducted. In \autoref{fig:kz_layout} two energy momentum cuts are shown along with the evolution of the \GM\ Fermi surface sheet with excitation energy.
One can see, that the $\alpha$- and $\beta$-band are showing \kz\ dispersion as both are subject to energy shifts with varying $h\nu$. Fitting of the dispersions obtained from MDCs shows that for $h\nu = \SI{70}{\eV}$ the top of the $\alpha$-band is at $\EB \sim \SI{7}{\milli\eV}$. For $h\nu = \SI{40}{\eV}$, the top of the $\alpha$-band is located at the Fermi level ($\EB = 0$) and for $h\nu = \SI{25}{\eV}$ it is at $\EB \sim \SI{-3}{\milli\eV}$, as shown in \autoref{fig:GammaPointBands}. Combining the spectra from a broader range of excitation energies, one obtains the \kz-dependence in \GM-Z-direction as seen in \autoref{fig:kz_layout}(c). Comparing these observations to the calculations, the \GM-point is identified with $h\nu \sim \SI{30}{\eV}$ and $h\nu \sim \SI{70}{\eV}$, where the width of the feature is minimal, while the maximum width denotes the Z-point. The $\alpha$-band thus creates a ``cigar shaped'' closed Fermi surface sheet around the Z-point of the brillouin zone. Similar features have been observed earlier and identified as a possible prerequisite for superconductivity in the iron-pnictides\cite{Borisenko2012}.The top of the $\beta$-band disperses from \SIrange{34}{15}{\milli\eV}. The $\gamma$ band shows no \kz-dispersion, as expected for the $d_{xy}$-state.
To quantify the size of the superconducting gap, which is supposed to be in the range of \SIrange{1.33}{2.2}{\milli\eV}\cite{Lin2011,Song2011}, we compared the experimental EDCs to EDCs obtained from modeled spectra\footnote{For the spectral function, the following model was used:\\$A(\boldk,\omega) = 2\pi[u_{\boldk}^2 \delta(\omega - E_{\boldk}) + v_{\boldk}^2 \delta(\omega + E_{\boldk})]$ with\\$u_{\boldk}^2 = \frac{1}{2} \left( 1 + \frac{\epsilon_{\boldk}}{E_{\boldk}}\right), v_{\boldk}^2 = \frac{1}{2} \left( 1 - \frac{\epsilon_{\boldk}}{E_{\boldk}}\right), E_{\boldk} = \sqrt{\epsilon_{\boldk}^2 + \Delta^2}$\\For further explanations see \cite{Evtushinsky2011}}.
 This comparison can be seen in \autoref{fig:kz_layout}(d). For \SI{1}{\kelvin}(blue) and \SI{11}{\kelvin}(red), the markers represent the experimental data while the solid lines show the EDCs from the modeled spectra. From fitting the model to the data, a maximum gap size of $\Delta_{\text{max}} = \SI{2}{\milli\eV}$ with a BCS-ratio of $2 \Delta/k_{\text{B}} = 5.7$ is estimated, similar to the findings for Fe\tsub{1.03}Te\tsub{0.7}Se\tsub{0.3}\cite{Nakayama2010}, which suggest a strong coupling scenario.\\
We have thus shown, that single crystalline FeSe, the structurally simplest member of the iron-pnictide/iron-chalcogenite family, also exhibits the simplest Fermi surface consisting of only one closed ``cigar shaped'' hole pocket around the Z-point of the Brillouin zone and a shallow, electron-like feature around the Brillouin zone corners (M-Point), propably consisting of two degenerate electron pockets. Due to the shallowness of the electron pockets it is hard to estimate the size of the corresponding Fermi surface sheets and thus to make a solid statement about the possibility of nesting.
The band structure is closer to the ones observed in different FeSe\tsub{$x$}Te\tsub{$1-x$} compounds than to the electronic structure of the recently investigated FeSe monolayers.
Temperature dependent measurements yielded a maximum size of the superconducting gap of about \SI{2}{\milli\eV} leading to a BCS ratio of 5.7. The d\tsub{xz}/d\tsub{yz} bands are renormalized by factors of 3 to 3.7, while the d\tsub{xy} band is renormalized by a factor of 9. This strongly differs from the conclusions in the previously mentioned studies\cite{Xia2009,Nakayama2010,Chen2010,Tamai2010}. We do not find a uniform renormalization, but a strongly orbital dependent one. But in contrast to Taimai et. al. we find comparable renormalization for the d\tsub{xz}/d\tsub{yz}, while the d\tsub{xy} shows a three times larger renormalization. In this aspect, the d\tsub{xy} band shows the most peculiar behaviour: it is not subject to \kz dispersion, shows a stronger renormalization than the other two bands, and is shifted to higher binding energies, thus not taking part in the formation of the Fermi surface. In addition to the renormalization, $k$-dependent shifts of the bandstructure of less than \SI{0.1}{\eV} lead to a dramatic change in the Fermi topology compared to DFT calculations, even in this structurally simplest iron-based superconductor.\\
While completing this manuscript, we noticed that the electronic structure of thick films as reported in recent studies\cite{Tan2013} is similar to what we observe in single crystals.

This work was supported by the DFG Schwerpunktprogramm 1458 (BO1912/3-1 and BO1912/2-2), Russian Foundation for Basic Research Grants 12-02-90405 and Russian Ministry of Science and Education Grant 8378.

%

\end{document}